\documentclass[prd,twocolumn,superscriptaddress,floatfix,amsmath,amssymb]{revtex4-1}

\usepackage[english]{babel}
\usepackage{hyperref}	
\usepackage{graphicx}
\usepackage{dcolumn}
\usepackage{bm}
\usepackage{hyperref}
\usepackage{braket}
\usepackage{amsmath,amssymb,latexsym}
\usepackage{cancel}
\usepackage{bbold}
\usepackage[usenames, dvipsnames]{color}
\usepackage{amsfonts}
\usepackage{comment}
\usepackage{natbib}
\usepackage{mathrsfs}
\usepackage{soul}
\newcommand{\spliteq}[1]{\begin{equation}
\begin{split}
#1
\end{split}
\end{equation}
}

\newcommand{\ii}{\mathrm{i}}

\begin{document}


\title{Asymptotically Limitless Quantum Energy Teleportation via Qudit Probes}

\author{Guillaume Verdon-Akzam}
 \affiliation{Department of Applied Mathematics, University of Waterloo, Waterloo, Ontario, N2L 3G1, Canada
}
 \affiliation{Institute for Quantum Computing, University of Waterloo, Waterloo, Ontario, N2L 3G1, Canada
}
\author{Eduardo Mart\'in-Mart\'inez}
 \affiliation{Department of Applied Mathematics, University of Waterloo, Waterloo, Ontario, N2L 3G1, Canada
}
 \affiliation{Institute for Quantum Computing, University of Waterloo, Waterloo, Ontario, N2L 3G1, Canada
}
\affiliation{Perimeter Institute for Theoretical Physics, 31 Caroline St N, Waterloo, Ontario, N2L 2Y5, Canada}

\author{Achim Kempf}
 \affiliation{Department of Applied Mathematics, University of Waterloo, Waterloo, Ontario, N2L 3G1, Canada
}
 \affiliation{Institute for Quantum Computing, University of Waterloo, Waterloo, Ontario, N2L 3G1, Canada
}
\affiliation{Perimeter Institute for Theoretical Physics, 31 Caroline St N, Waterloo, Ontario, N2L 2Y5, Canada}

\date{\today}

\begin{abstract}
We propose a modified Quantum Energy Teleportation (QET) scheme that uses arbitrary-dimensional qudit probes and polynomially localized Hamiltonians. We find that with an appropriate scaling of parameters, the teleported energy scales with the teleportation distance more favourably than the non-local tails of the Hamiltonians. We show that by allowing the exchange of arbitrary amounts of information between agents and in a suitable limit, an arbitrarily large amount of energy can be teleported through a massless quantum field. 

\end{abstract}

\maketitle


\section{Introduction} 
Quantum Energy Teleportation (QET) \cite{Hotta2008, Hotta2010, Hotta2010b, Hotta2014} is a protocol that allows the transmission of energy between a sender, A, and a receiver, B, without energy propagating through the field from A to B. 

Instead, the protocol relies on pre-existing field entanglement and a classical communication channel between A and B. First, A invests energy to measure the field at her location and then informs B via a classical channel of the outcome. Then, B can to some extent predict an upcoming field fluctuation at his location and extract work from it. It is known that the energy invested by A on average exceeds the amount of work extracted by B. Here, we modify the conventional QET protocol, so that A is able, at increased energetic expense, to extract more information from her field measurement and communicate it (classically or quantumly) to B. We show that this can indeed enable B, in principle, to extract an arbitrarily large amount of work from the field, again without energy having travelled from A to B.

In the conventional QET protocol \cite{Hotta2010}, the sender, A, couples a spatially localized two-level quantum system (also called an Unruh-DeWitt (UdW) detector) \cite{DeWitt}) to the field for an instant in time, thereby in effect probing the vacuum fluctuations of the field. A subsequent projective measurement of this qubit system yields a single classical bit, which is transmitted to the receiver, B, via classical communication. Depending on that bit, B then performs one of two localized operations by coupling another localized system to the field. On average, A will invest energy into the field at the moment of her measurement, while B will extract work via his action on the field. In this protocol, no energy flows from A through the field to B. QET is, therefore, a method for overcoming the constraints of strong local passivity \cite{Frey2014} of Gibbs states (namely no local energy extraction) such as the vacuum, namely by supplementing local operations with classical communication (LOCC). Experimental realizations of QET are being considered, with  proposals to realize QET, e.g., in quantum Hall systems \cite{Yusa2011}, or with trapped ions \cite{Hotta2009}.

In regards to resources, QET exploits the pre-existing spatial correlations in the field fluctuations which, in the case of quantum fields in the vacuum, originate in the vacuum entanglement. It has also been determined for the cases of minimal (two-qubit) and linear harmonic chain QET \cite{Hotta2010a, Nambu2010} that it is the consumption of mutual information between A and B which allows for the teleportation of energy between the two parties. 

The question naturally arises then about scenarios where A probes the field with a higher-dimensional quantum system and then transmits the outcome as multiple bits or even qubits to B. More correlations might be harvested in this way from the field \cite{Pozas-Kerstjens:2015} and could this then allow B to extract more energy from the field, and if so, is there an upper bound?
To explore this idea, we generalize the QET protocol by allowing A to probe the field by locally coupling to it a quantum system of arbitrary finite dimensions, i.e., a $d$-dimensional \textit{qudit}. We also allow A to transmit the resulting information to B as bits or qubits. In the latter case, A transmits her entire qudit to B.

We will now show that if A and B use qudits with  suitably-chosen interaction Hamiltonians with the field, then this indeed yields a greater efficiency of teleportation of energy, i.e., an increased ratio of energy extracted versus invested. 

\subsection{Mathematical Setup} 
\subsubsection{Qudits}
We begin with the description of the qudits. To this end, let us consider the set of unitary, traceless and non-Hermitian (when $d>2$) generalized Pauli matrices, often called the Weyl (or Heisenberg-Weyl) operators \cite{Bertlmann2008, Vourdas2004}, which act on the Hilbert space $\mathscr{H}=\mathbb{C}^d$. 
The generalized Pauli $Z$ matrix, also called the \textit{clock} matrix, is defined through
\begin{equation}
Z =  \sum_{j=0}^{d-1}e^{2\pi \ii j/d}\ket{z_j}\!\bra{z_j},
\end{equation}
where $\{\ket{z_j}|\,j\in\mathbb{Z}_d\}$ is the standard basis of $\mathbb{C}^d$ (the eigenbasis of $Z$). The generalized Pauli $X$ matrix, or \textit{shift} matrix, is defined as
\begin{equation}
X = \sum_{j=0}^{d-1}\ket{z_{j+1\,(\text{mod}\ d)}}\!\bra{z_j}.
\end{equation} Note that both matrices share the same spectrum which consists of all the $d^\text{th}$ roots of unity, $\{e^{-2\pi \ii j/d}| j \in \mathbb{Z}_d\}$.  

The clock and shift {matrices} can be viewed as displacement operators on a toroidal lattice \cite{Vourdas2004} whose set of points, $\mathcal{W}\equiv\{e^{-\pi \ii ab/d}Z^aX^z_b|\, a,b \in \mathbb{Z}_d\}$, forms the Heisenberg-Weyl group.   
Also note that when $d$ is a prime number, the $Z$ eigenbasis, and the $XZ^m,\ m=0,..., d-1$ eigenbases are all mutually unbiased \cite{Bandyopadhyay2001}. Finally, the most important feature of these matrices for our purposes will be that they obey the following \textit{braiding relation} 
\begin{equation}\label{eq:braiding}
ZX = e^{2\pi \ii /d}XZ.
\end{equation}
  
\subsubsection{Quantum Field}
For the quantum field, we consider a 1+1 dimensional massless Klein-Gordon field, quantized in the {null} coordinates ($x^\pm \equiv t\pm x$), we consider the left and right-moving sectors independently \cite{Hotta2008}:
\spliteq{
\Phi(x) = \Phi_+(x^+)+ \Phi_-(x^-),
}
where
\begin{align}
\Phi_\pm(x^\pm)
&\equiv  \int_{\mathbb{R}^+}\!\!\!\!\! \text{d}\omega\tfrac{1}{\sqrt{4\pi\omega}} \left[a_{\pm\omega} e^{-\ii\omega x^\pm}+a^{\dagger}_{\pm\omega} e^{\ii\omega x^\pm}\right]. 
\end{align}  The canonically conjugate momenta of these field components are given by $\Pi_\pm(x^\pm)\equiv \partial_\pm\Phi_\pm(x^\pm)$, or explicitly
\spliteq{
\Pi_\pm(x^\pm)=  -\ii\int_{\mathbb{R}^+}\!\!\!\!\! \text{d}\omega\sqrt{\tfrac{\omega}{4\pi}} \left[a_{\pm\omega} e^{-\ii\omega x^\pm}-a^{\dagger}_{\pm\omega} e^{\ii\omega x^\pm}\right].
}
The Hamiltonian density of this field is given by 
\spliteq{
\mathcal{H}_F (x)&\equiv \tfrac{1}{2} : \Pi_+(x)^2: +\tfrac{1}{2} : \Pi_-(x)^2:\\
&\equiv \varepsilon_+(x) +\varepsilon_-(x)
}
where $:*:$ denotes normal ordering and the labels $+,-$ denote left and right-moving components respectively.
 Explicitly, the left-moving modes' energy density is given by
\spliteq{
&\varepsilon_+(x) =\!\!\int_{(\mathbb{R}^+)^2}\!\!\!\!\!\!\!\!\!
\text{d}\omega \text{d}\omega' \frac{ \sqrt{\omega\omega'}}{8\pi} \Big[2a^{\dagger}_{+\omega} a_{+\omega'} e^{\ii(\omega-\omega') x} \\&\qquad -a^{\dagger}_{+\omega} a^{\dagger}_{+\omega'} e^{\ii(\omega+\omega') x}-a_{+\omega} a_{+\omega'} e^{-\ii(\omega+\omega') x}\Big].
}
The free Hamiltonian of this field is given by a spatial integral of the above Hamiltonian density
\spliteq{
H_F 
&= \int_{\mathbb{R}}\!\! \text{d}x\, \mathcal{H}_F(x) = \int_{\mathbb{R}}\!\! \text{d}x \left[\varepsilon_+(x^+)+\varepsilon_-(x^-)\right]\\
&= \int_{\mathbb{R}^+}\!\!\!\!\! \text{d}\omega\ \frac{\omega}{2} \left[a^{\dagger}_{+\omega} a_{+\omega}+a^{\dagger}_{-\omega} a_{-\omega}\right],
}

Since the left moving ($+$) and right-moving ($-$) parts of the field are decoupled, we can consider acting upon each individually.  
In order to make it plain that the energy will be teleported rather than transported from A to B, we will couple A and B only to the left moving modes, while B is allowed to be to the right of A. As we will work exclusively with left moving modes, for convenience we drop their $+$ subscript.

\section{The New Protocol} 
\subsection{Description}
We first describe the scenario where A and B use localized operations and quantum communication in order to teleport energy. 

\subsubsection{Sender's Action on the Field}
We assume that the field begins in the vacuum state $\ket{\bm{0}}$, hence the expectation of the renormalized stress energy tensor vanishes everywhere. At $t=0^-$, A is in an arbitrary initial state $\ket{\mathrm{A}_0} \in \mathscr{H}_A \equiv \mathbb{C}^d$. To perform QET, A's qudit is then at $t=0$ coupled for an instant to the field via an interaction Hamiltonian $\delta(t) H_\mathcal{A}$, which is a spatial smearing of a Hamiltonian density with respect to a compactly supported smearing function $\lambda:\mathbb{R}\rightarrow\mathbb{R}$, i.e., $H_\mathcal{A} \equiv \int_{\mathbb{R}} \text{d}x\ \lambda(x) \mathcal{H}_\mathcal{A}(x)$, where
\spliteq{\label{eq:A_int_dens}
\!\!\!\!\mathcal{H}_\mathcal{A}(x) &\equiv -\ii\! \int_{\mathbb{R}^+}\!\!\!\!\! \text{d}\omega \sqrt{\tfrac{\omega}{4\pi}} \left(X \otimes a_\omega e^{-\ii\omega x}\! -\! X^\dagger \otimes a^\dagger_\omega e^{\ii\omega x}\right)\!\!
} becomes the Hamiltonian density when multiplied by $\lambda(x)$. Later on, it will be useful for us to write the spatial coupling factor as the product of the coupling strength times a  spatial profile (with dimensions of inverse length) of unit  $L^1(\mathbb{R})$ norm, i.e. 
\begin{equation}\label{eq:factor_A}
\lambda(x) =\lambda_0 F_\textsc{a}(x) 
\end{equation}
where
\begin{equation}\lVert F_\textsc{a}\rVert_1\equiv\int_{-\infty}^\infty\!\!\!\! \text{d}x\, F(x)=1\end{equation}

 Note that in the $d=2$ qubit case, $X=X^\dagger$ and we simply obtain $\mathcal{H}_\mathcal{A}(x)  = X\otimes \Pi_+(x)$, which is the Unruh-DeWitt (UdW) detector model \cite{DeWitt}, similar to conventional QET \cite{Hotta2010}. Notice that for $d>2$, $X\neq X^\dagger$ and therefore the Hamiltonian \eqref{eq:A_int_dens} cannot be written as a tensor product of a qudit observable and a field observable. In fact, the form of  \eqref{eq:A_int_dens} resembles a counter-rotating wave interaction Hamiltonian, as seen in the context of quantum optics. The subtleties regarding the locality of such an interaction model (pointed out in \cite{EMM2015}), will be discussed in full detail in section \ref{sec:asym_loc}.

Using the eigendecomposition of the $X$ operator, $X= \sum_{j} \Upsilon^j\ket{x_j}\!\bra{x_j}$ {(where $\Upsilon \equiv e^{2\pi i /d}$)}, and Fourier transforming the spatial profile, $\tilde{\lambda}(\omega) \equiv  \int_{\mathbb{R}} \text{d}x~e^{\ii\omega x}\lambda(x)$, we obtain
\spliteq{\label{eq:H_A-final}
H_\mathcal{A} &= \ii  \sum_{j=0}^{d-1}\ket{x_j}\!\bra{x_j} \otimes\int_{\mathbb{R}^+}\!\!\!\!\! \text{d}\omega \left(\zeta_{j\omega} a^\dagger_\omega  - \zeta^*_{j\omega} a_\omega \right),
}
where $\zeta_{ j\omega } \equiv \Upsilon^{-j} \alpha_\omega,\ \alpha_\omega =\sqrt{\tfrac{\omega}{4\pi}}\tilde{\lambda}(\omega) \in \mathbb{C}$.
In effect, $H_\mathcal{A}$ is a multi-mode generator of vacuum displacement where the amounts of displacement depend the state of the probe. This is similar to a controlled quantum gate, the control being our qudit probe in this case. The time evolution operator generated by the above interaction Hamiltonian reads
\spliteq{\label{eq:U_A}
U_{\mathcal{A}} &\equiv \text{T~exp}\left(-\ii \!{\int_{\mathbb{R}}\!\! \text{d}t\, \delta(t)H_{\mathcal{A}} }\right) = \exp(-\ii H_{\mathcal{A}})\\
&= \sum_{j=0}^{d-1} \ket{x_j}\!\bra{x_j} \otimes D(\bm{\zeta}_j)
}
where $D(\bm{\zeta}_j) \equiv \exp\left(\int_{\mathbb{R}^+}\text{d}\omega \left(\zeta_{j\omega} a^\dagger_\omega  - \zeta^*_{j\omega} a_\omega \right)\right)$ is a multi-frequency displacement operator acting on the Fock space of the field $\mathscr{F}$. 
Given the initial state $\ket{\Psi_0}\equiv\ket{\mathrm{A}_0}\otimes \ket{\bm{0}}$, we can compute the state after A's interaction,
\spliteq{\label{eq:A_state}
\ket{\Psi_1} \equiv U_\mathcal{A} \ket{\Psi_0}  = \sum_{j=0}^{d-1} \braket{x_j|\mathrm{A}_0}\ket{x_j} \otimes \ket{\bm{\zeta}_j},
}
where $\ket{\bm{\zeta}_j} \equiv \bigotimes_{\omega=0}^\infty \ket{\zeta_{j\omega}} =\bigotimes_{\omega=0}^\infty \ket{e^{-2\pi \ii j/d}\alpha_{\omega}} $, the boldface is used to indicate that it is a multi-frequency coherent state. 
The expectation value of the energy $E_A \equiv \bra{\Psi_1}I\otimes H_F \ket{\Psi_1}$ of this state reads
\begin{equation}\label{eq:E_A}
\!\!\!E_A\! = \! \frac{1}{2}\!\int_{\mathbb{R}^+}\!\!\!\!\! \text{d}\omega\,\omega  |\alpha_{\omega}|^2 \!=\! \bra{\bm\alpha}H_F\ket{\bm\alpha} \!=\!\frac{1}{8}\! \int_{\mathbb{R}}\!\! \text{d}x\, (\lambda'(x))^2, \end{equation}

where the $\lambda'(x) \equiv \partial_x \lambda(x)$, appeared from applying an inverse Fourier transformation (proof of the above can be found in appendix \ref{ApA}). We find that it is the same as that of a single coherent state, whose energy depends on the $2$-norm of the derivative of the spatial coupling $\lambda$. 
{Note that we assume A's qudit to have no free Hamiltonian, i.e., that it is a gapless qudit.}

\subsubsection{Receiver's Action on the Field}
A now sends her `measurement' information to B. To do so, she sends her probe qudit directly to B via a quantum channel. We assume that the quantum channel preserves the qudit as is, whereas the field evolves according to its free Hamiltonian during the time of qudit transmission, $T$. The state of the field and A's qudit after this time evolution is given by
\begin{equation}\label{eq:after_transfer}
\ket{\Psi_2} \equiv U_\mathcal{F}(T)\ket{\Psi_1}=  \sum_{j=0}^{d-1} \braket{x_j|\mathrm{A}_0}\ket{x_j} \otimes \ket{\bm{\Lambda}_j}
\end{equation}
where $U_\mathcal{F}(T) \equiv (I\otimes e^{-\ii H_FT})$,  $\ket{\bm{\Lambda}_j} \equiv \bigotimes_{\omega=0}^\infty \ket{\Lambda_{j\omega}}$, and $\Lambda_{j \omega}\equiv e^{-\ii\omega T} \zeta_{ j\omega }$. Note that by energy conservation in the field during free evolution, $E_A =\bra{\Psi_2}I\otimes H_F\ket{\Psi_2}$.

Now, at time $T$, B couples the probe he received from A with the field according to an interaction Hamiltonian $\delta(t) H_\mathcal{B}$, similar to A's. It has again a spatial smearing of a Hamiltonian density $H_\mathcal{B} \equiv \int_{\mathbb{R}} \text{d}x\ \mu(x) \mathcal{H}_\mathcal{B}(x)$, with respect to a compactly supported function, $\mu:\mathbb{R}\rightarrow\mathbb{R}$, where
\spliteq{\label{eq:B_int_dens}
\!\!\!\!\mathcal{H}_\mathcal{B}(x) &\equiv  \int_{\mathbb{R}^+}\!\!\! \text{d}\omega \tfrac{1}{\sqrt{4\pi\omega}} \left(Z \otimes a_\omega e^{-\ii\omega x}\! +\! Z^\dagger \otimes a^\dagger_\omega e^{\ii\omega x}\right)\!.\!\!\!
}
Notice the difference with (\ref{eq:A_int_dens}), here we use $Z$ instead of $X$ and $\mu(x)$ instead of $\lambda(x)$, otherwise, \eqref{eq:B_int_dens} is to $\Phi_+(x)$ what \eqref{eq:A_int_dens} is to $\Pi_+(x)$; in the $d=2$ case, the above becomes $Z\otimes \Phi_+(x)$. Same as in \eqref{eq:factor_A}, it will be useful to write $\mu(x)$ as the product of a coupling strength and a smearing function  (with dimensions of inverse length) of unit $L^1(\mathbb{R})$ norm: $\mu(x)\equiv \mu_0F_\textsc{b}(x)$, where $\lVert F_\textsc{b}\rVert_1 =1$. Note that because A and B's couplings are of different nature (in the limit $d=2$ they constitute respectively field and momentum UdW coupling), the coupling strengths $\mu_0$ and $\lambda_0$ have differing dimensions, as is clear also from the different powers of $\omega$ in \eqref{eq:A_int_dens} and \eqref{eq:B_int_dens}. Similarly to (\ref{eq:U_A}), the time evolution operator associated to B's interaction is a controlled field displacement, given by \spliteq{\label{eq:B_U_B}
\!\!\!U_{\mathcal{B}} \equiv \text{Texp}\left(-\ii\! {\int_{\mathbb{R}}\!\! \text{d}t\, \delta(t)H_{\mathcal{B}}} \right) =  \sum_{i=0}^{d-1} \ket{z_i}\!\bra{z_i} \otimes D(\bm{\xi}_i),
}
where $D(\bm{\xi}_i) \equiv \exp\left(\int_{\mathbb{R}^+}\text{d}\omega \left(\xi_{i\omega} a^\dagger_\omega  - \xi^*_{i\omega} a_\omega \right)\right)$, $\xi_{ i\omega } \equiv \Upsilon^{-i} \beta_\omega$, $\beta_\omega \equiv\tfrac{-i}{\sqrt{4\pi\omega}}\tilde{\mu}(\omega)$. 
The final state of the protocol, $\ket{\Psi_3} \equiv U_\mathcal{B}\ket{\Psi_2}= U_\mathcal{B}U_\mathcal{F}(T)U_\mathcal{A}\ket{\Psi_0}$, immediately after B's interaction, is given by
\spliteq{\label{eq:psi_3}
\ket{\Psi_3}=\sum_{i,j=0}^{d-1} \left(\braket{z_i|x_j}\!\braket{x_j|\mathrm{A}_0}\right)\ket{z_i} \otimes D(\bm{\xi}_i)\ket{\bm{\Lambda}_j}.
}

\subsection{Energy Extraction}
To show that B can extract {an average positive net} work from the field through his instantaneous local interaction, we compute the expectation value of the energy of the field in the final state, and compare it to the energy right before B's operation,
\begin{eqnarray}\label{eq:Delta_E}
\!\!\!\!\!\!\!&\!\!\!\!\!\!\!\!\!\!\!\!\!\!\!\!\!\!\!\Delta E \equiv \bra{\Psi_3}I\otimes H_F\ket{\Psi_3} - E_A = \tfrac{1}{2}\textstyle{\int_{\mathbb{R}^+}} \text{d}\omega\ \omega \  |\beta_\omega|^2\  \\&+ \text{Re}\left[\left(\textstyle{\int_{\mathbb{R}^+}} \text{d}\omega\ \omega e^{\ii\omega T}\alpha^*_{\omega}\beta_{\omega}\right)\bra{\mathrm{A}_0}X Z^\dagger\ket{\mathrm{A}_0}e^{(\Upsilon-1)\lVert\bm{\alpha}\rVert^2}\right]\nonumber
\end{eqnarray} where 
\begin{equation}
\lVert\bm{\alpha}\rVert^2 \equiv \textstyle{\int_{\mathbb{R}^+}\! \text{d}\omega\  |\alpha_\omega|^2 = \int_{\mathbb{R}^+}\! \text{d}\omega\  \tfrac{\omega}{4\pi}|\tilde{\lambda}(\omega)|^2}.
\end{equation}
Note that the key property used in the computation of the above was the braiding relation \eqref{eq:braiding}, a proof of the above is provided in appendix \ref{ApA}. Since this $\Delta E$ is the change in the energy in the field caused by B's local interaction, then any energy gained/lost in the field is necessarily lost or gained by B, i.e., by the classical agent which drives the switching of the interaction Hamiltonian \eqref{eq:B_int_dens}. 

\subsubsection{Classical Communication}
So far, we assumed that A and B used a quantum channel for communication, i.e., that A sends her qudit to B.  
Here we outline how an equivalent classical communication version of the protocol can be obtained. After A's interaction (\ref{eq:A_state}), A performs a projective measurement in the $Z$ basis (or more generally in the qudit's eigenbasis for B's interaction Hamiltonian). Given an observed outcome $i$, the (pre-normalized) state of the field is then given by
\begin{equation}
\frac{\ket{\phi_i}}{\sqrt{d}}\equiv  (\bra{z_i}\otimes I)\ket{\Psi_1}= \sum_{j=0}^{d-1} \braket{x_j|\mathrm{A}_0}\braket{z_i|x_j}  \ket{\bm{\zeta}_j},
\end{equation}
where the $1/\sqrt{d}$ factor to normalize $\ket{\phi_i}$ comes from the fact that $Z$ and $X$ are mutually unbiased bases \cite{Bandyopadhyay2001}. 
Following this projective measurement, A then would transmit this measurement result; the value $i\in \mathbb{Z}_d$, to B via a $\log_2(d)$ bit string. After a free evolution of the field for a time $T$, B then applies the $D(\bm{\xi}_i)$ unitary  (as defined in  (\ref{eq:B_U_B})), dependent on the string received. The state of the field then becomes
\begin{equation}
\frac{\ket{\phi'_i}}{\sqrt{d}} \!\equiv\! \left(\bra{z_i}\otimes D(\bm{\xi}_i)\right)\ket{\Psi_2} \!= \!\sum_{j=0}^{d-1} \braket{x_j|\mathrm{A}_0}\!\braket{z_i|x_j}  D(\bm{\xi}_i)\ket{\bm{\Lambda}_j},
\end{equation}
one can notice the similarity with \eqref{eq:psi_3}; indeed $\ket{\phi'_i} = \sqrt{d}(\bra{z_i}\otimes I)\ket{\Psi_3}$.
Hence, taking an expectation value of $H_F$ for the above state while averaging over the possible values of $i$ ($d$ measurement outcomes) yields the first term from (\ref{eq:Delta_E}),
\spliteq{
\bra{\Psi_3} I\otimes H_F \ket{\Psi_3}  =\frac{1}{d}\sum_{i=0}^{d-1}  \bra{\phi'_i}H_F \ket{\phi'_i}.
}
 We conclude that the quantum and classical communication versions of the protocols yield the same outcome in terms of average teleported energy.

\begin{figure}[h]
 \begin{center}
\includegraphics[width=1\columnwidth]{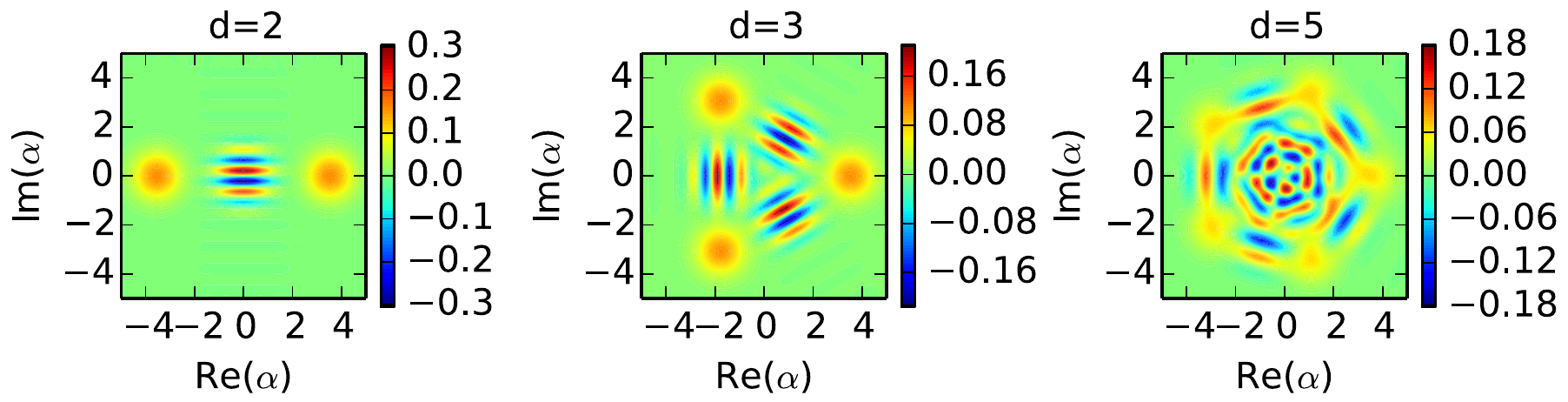}
\caption{Single-mode Wigner pseudoprobability distribution for the cat state $(\bra{0}\otimes I)\ket{\Psi_1}$, i.e., the phase space representation of a single mode of the field if A measured her qudit in the $\ket{0}$ state after her interaction. The initial qudit state $\ket{\mathrm{A}_0}$ chosen here was an $XZ^\dagger$ eigenstate. The parameter $\alpha =2.5$ is fixed and we plot for various probe dimensions, labelled $d$. As the dimensions grow, the distribution loses some of its isotropy, allowing for energy extraction when displaced along an appropriate direction.
} \label{fig:catstates}
\end{center}
\end{figure}

For illustration, in the classical communication version of the new protocol, let us assume A begins with an initial qudit state $\ket{\mathrm{A}_0}$ that is a basis state mutually unbiased to the $X$ and $Z$ eigenbases (e.g. an $XZ^m$ eigenstate for $d$ prime, $m\in\mathbb{Z}_d$). Then after A's interaction through (\ref{eq:A_int_dens}) and projective measurement on the qudit, the field is in a uniform superposition of $d$ coherent states in the field, i.e., a Cat state, akin to those depicted in Fig.\ref{fig:catstates}. 

\subsubsection{Setup for Optimal Extraction}
Let us now analyse in what circumstances B was able to extract work from the field, that is, $\Delta E<0$. We notice that the first term in (\ref{eq:Delta_E}) is positive and is the energy that B invests in the field by switching on his coupling, similar to the energy A invested in (\ref{eq:E_A}). 

The second term of (\ref{eq:Delta_E}) is the one of interest, since it may become negative and overcome the positivity of the first term. The Fourier integral, $\mathcal{I} \equiv\textstyle{\int_{\mathbb{R}^+}} \text{d}\omega\ \omega e^{\ii\omega T}\alpha^*_{\omega}\beta_{\omega}$, using \cite{Messiah1961}, can be rewritten as 
\begin{equation}
\!\!\mathcal{I}\! =\!-\frac{1}{4}\! \int_{\mathbb{R}}\!\! \text{d}x\,\lambda(x)\mu'(x\!-\!T) \!+\! \frac{\ii}{4\pi}\! \int_{\mathbb{R}^2}\!\!\!\! \text{d}x\, \text{d}y\,\frac{\lambda(x)\mu(y)}{(y\!-\!x\!+\!T)^2}.
\end{equation}
We see that the real part is non-vanishing strictly on the left-moving light-front, while the imaginary part is non-vanishing everywhere.
Hence, only the imaginary part can allow B to extract work even if he is not light-like separated and to the left of A; this is the term that gives us the ``teleportation'' aspect of QET. Notice that the second term in (\ref{eq:Delta_E}) is comprised of the real part of the product $\mathcal{I}\cdot \Gamma$ where
\begin{equation}
\Gamma\equiv\bra{\mathrm{A}_0}X Z^\dagger\ket{\mathrm{A}_0}e^{(e^{2\pi\ii /d}-1)\lVert\bm{\alpha}\rVert^2}
\end{equation} is a functional dependent on the coupling profile and strength of A, the dimensions of the probe, and the initial qudit state. In general, we will have $|\Gamma|\leq 1$.  Looking back at (\ref{eq:Delta_E}), one sees that to maximize the teleported energy, the imaginary part of $\Gamma$ needs to be as large as possible.

To this end, we note that the spectrum of the unitary $XZ^\dagger$ is $\text{spec}(XZ^\dagger)=\{-e^{\pi \ii(2j+1)/d}| j\in \mathbb{Z}_d\}\subset\mathbb{C}$. The convex hull, $S$, of this spectrum is a regular $d$-gon inscribed in the unit circle, and in the limit $d\rightarrow \infty$, $S$ converges to the unit disk.
For any point $s\in S$, there is 
a state $\ket{\mathrm{A}_0}$ such that $\bra{\mathrm{A}_0}XZ^\dagger\ket{\mathrm{A}_0} =s$. This is because, decomposing $\ket{\mathrm{A}_0}$ into $XZ^\dagger$'s eigenbasis, $\bra{\mathrm{A}_0}XZ^\dagger\ket{\mathrm{A}_0}$ yields the convex combinations of the eigenvalues.
Therefore, for any $\chi \equiv e^{(\Upsilon-1)\lVert\bm{\alpha}\rVert^2}$ we can choose a  $\ket{\mathrm{A}_0}$ such that 
\begin{equation}\label{eq:Gamma_bound}
|\text{Im}(\Gamma)| = |\text{Im}[\chi \bra{\mathrm{A}_0}XZ^\dagger\ket{\mathrm{A}_0}]| \geq \cos(\tfrac{\pi}{d})|\chi|.
\end{equation}

\noindent Now assuming B is off of A's left-moving lightfront, the total work extracted by B, in terms of both A and B's spatial couplings, is given by 
\spliteq{\label{eq:delta_E_spatial}\!\!\!\!\Delta E=\frac{\mu_0^2}{8}\! \!\int_{\mathbb{R}}\!\! \text{d}x (F_\textsc{b}(x))^2
\!-\!\text{Im}(\Gamma)\frac{\lambda_0\mu_0}{4\pi}\!\! \int_{\mathbb{R}^2}\!\!\!\!\text{d}x\, \text{d}y\frac{F_\textsc{a}(x)F_\textsc{b}(y)}{(y\!-\!x\!+\!T)^2},\!\!\!\!
}
where we used that $\lambda(x)=\lambda_0 F_\textsc{a}(x)$, $\mu(x)=\mu_0 F_\textsc{b}(x)$ as defined above. There are two ways to make the second term dominate: 
one can scale down  $\mu_0$, as in conventional QET \cite{Hotta2010}, but this eventually decreases the amount of energy that can possibly be teleported. Let us consider the remaining option, to scale up $\lambda_0$. Since $\Gamma$ depends on $\lambda_0$ and $d$, we must optimize the scaling of both variables. 

\subsection{Dimensions vs. Energy Scaling}\label{sec:dim_vs_E}
Suppose A linearly increases the strength of her interaction in a spatially uniform manner i.e., for $\theta\in \mathbb{R^+}$, $\lambda_0\mapsto \theta \lambda_0$, $F_\textsc{a}(x)\mapsto F_\textsc{a}(x)$ (recall $\lambda(x) \equiv \lambda_0 F_\textsc{a}(x),\ \lambda_0\equiv \lVert\lambda\rVert_1$), causing $\lVert\bm{\alpha}\rVert^2 \mapsto \theta^2\lVert\bm{\alpha}\rVert^2 $. Meanwhile, suppose that A also superlinearly increases the dimensions of the measurement probe $d\mapsto \lceil\theta^{1+\epsilon}\rceil d$, for some $\epsilon>0$. 
Using the bound in (\ref{eq:Gamma_bound}), we obtain\vspace{-2.5mm}
\begin{equation}
\!\!\!\!\left|\text{Im}(\Gamma)\right|  \geq \cos(\tfrac{\pi}{d\lceil\theta^{1+\epsilon}\rceil})e^{\left(\cos\left(\tfrac{2\pi}{\lceil\theta^{1+\epsilon}\rceil d}\right)-1\right)\theta^2\lVert\bm{\alpha}\rVert^2} .
\end{equation}
In the large $\theta$ limit, we have  
\begin{equation}\label{eq:lrg_theta}
\left|\text{Im}(\Gamma)\right| \geq1-\tfrac{2\pi^2\lVert\bm{\alpha}\rVert^2}{\theta^{2\epsilon} d^2} + \mathcal{O}(\theta^{-2-2\epsilon}) \overset{\theta\rightarrow\infty}{\longrightarrow} 1.
\end{equation}
Finally, from (\ref{eq:delta_E_spatial}), we see that the extracted energy is then linearly increasing with the scaling, $\Delta E \sim -\theta$. Note that a linear scaling of the dimensions ($\epsilon =0$) suffices,  but in that case $\left|\text{Im}(\Gamma)\right|\gtrapprox e^{-2\pi^2\lVert\bm{\alpha}\rVert^2/d^2}$ as $\theta \rightarrow\infty$.

We conclude from the above protocol that there so far seems to be no upper bound to the total energy recoverable by B, independently of the shape or spread of his spatial coupling. This is in contrast to the $d=2$ case in previous descriptions of QET \cite{Hotta2010}. The main limiting factor in the $d=2$ case, where we have $|\chi| =e^{-2\lVert\bm{\alpha}\rVert^2} $, is the exponential decay with respect to the number expectation value $\lVert\bm{\alpha}\rVert^2= \int_0^\infty \text{d}\omega\,|\alpha_\omega|^2\sim\lambda_0^2$.
Our new protocol circumvents this problem by countering the exponential decay via scaling of the probe dimensions. We further elaborate on this advantage over traditional UdW-type detectors below.

\section{Discussion} \subsection{Advantage Over UdW} In order to obtain the favorable energy scaling, the choice of interaction Hamiltonians in \eqref{eq:A_int_dens} and \eqref{eq:B_int_dens} was important. It would not have sufficed to couple the qudits as conventional UdW detectors, through  $\mathcal{H}\propto J_x\otimes\Phi$ where $J_x$ is a generator of a $d$-dimensional representation of SU(2) \cite{DeWitt}. 
In that case, when computing the second term of (\ref{eq:Delta_E}) there would be a sum of terms which are of the form \spliteq{\sim\braket{m\bm{\alpha}|k\bm{\alpha}}=e^{\left(mk-(m^2+k^2)/2\right)\lVert\bm{\alpha}\rVert^2},}
where $m,k\in\{-\tfrac{d}{2},\dots,\tfrac{d}{2}\}$ or $\{-\tfrac{d-1}{2},\dots,\tfrac{d-1}{2}\}$ for $d$ even or odd respectively. This means that the teleportation term in \eqref{eq:Delta_E}  would be  exponentially suppressed with respect to the amplitude of the coherent state generated by A's action on the field. Hence, for the UdW-like case, scaling A's coupling strength will not help QET.

In contrast, the Hamiltonian \eqref{eq:A_int_dens} yields contributions to the teleportation term which are of the form \spliteq{\label{eq:inner_prod}
\!\!\!\sim\braket{\bm{\Lambda}_m|\bm{\Lambda}_k} = \braket{\bm{\Upsilon}^{-m}\bm{\alpha}|\bm{\Upsilon}^{-k}\bm{\alpha}} = e^{(e^{2\pi\ii (m-k)/d}-1)\lVert\bm{\alpha}\rVert^2}\!\!\!
} where $m,k\in\mathbb{Z}_d$. The combination of these complex terms, which then form $\chi = \exp\left((e^{2\pi\ii /d}-1)\lVert\bm{\alpha}\rVert^2\right)$ in \eqref{eq:Delta_E}, causes the teleportation term to scale with $d$ and $\lVert\bm{\alpha}\rVert^2$ in the manner discussed above, overcoming the previous limitations.

\subsection{Asymptotic Locality}\label{sec:asym_loc} 

Flanagan's quantum inequalities \cite{Flanagan97} establish a finite upper bound for any QET protocol over the field vacuum where A and B use strictly local operations \cite{Hotta2014}. However, the protocol proposed in this paper may seem to overcome the limitations of such bound. The reason for this is related to the degree of localization of A and B's Hamiltonians \eqref{eq:A_int_dens} and \eqref{eq:B_U_B}.

As mentioned previously, we will see in this section that A and B's interaction Hamiltonians are not strictly local. This may seem problematic given that in order to be able to draw the conclusion that \it all \rm energy extracted by B has been teleported one would have to assume that A and B only use strictly local operations. While indeed, generally not all of the energy that B extracts will have been teleported, we will now show that in a suitable scaling limit the interaction Hamiltonians become approximately local (to an arbitrary degree) when the distance between A and B increases, while still allowing for limitless quantum energy teleportation.

A first way of characterizing the non-locality of our Hamiltonian is to express it in terms of the field operator and its canonical conjugate momentum. To do that we first decompose the qudit's Weyl operators into Hermitian and anti-Hermitian parts, let $C=(X+X^\dagger)/2$, $S=\ii(X-X^\dagger)/2$, then we can rewrite \eqref{eq:A_int_dens} as
\spliteq{
\!\!\mathcal{H}_\mathcal{A}(x)=\frac{C}{2}\Pi_+(x) \!+\! \frac{S}{8\pi}\! \int_{\mathbb{R}}\! \text{d}y \left( \frac{\Phi_+(y)}{(x-y)^2} \!+\!\frac{\Pi_+(y)}{(x-y)} \right).
}
We see that the anti-Hermitian component corresponds to a power-law decay smearing of local observables. Notice that the non-local term vanishes in Hotta's original QET protocol \cite{Hotta2010} i.e., when $d=2$ (the Hamiltonian is simply the usual Unruh-DeWitt Hamiltonian).

As a consequence, to label the proposed protocol as legitimate QET with such non-local interaction, it is necessary that, as the distance between A and B increases, the non-local contribution to the field energy due to A's field interaction  decays faster than the amount of teleported energy retrieved by B from the field. This would recover a clean QET protocol in the asymptotic limit of large teleportation distances.

To study this, we must compute the energy density in the field after by A's interaction, that is, \mbox{$E_A(x) \equiv \bra{\Psi_1}I\otimes \varepsilon_+(x) \ket{\Psi_1}$}, giving
\spliteq{\label{eq:E_dens_A}
E_A(x) = &\frac{1}{16}(1+\text{Re} \braket{X^2})(\lambda'(x))^2 \\ &+\frac{1}{16\pi^2}(1-\text{Re} \braket{X^2})\left({\int_{\mathbb{R}}\text{d}y\, \frac{\lambda(y) }{(x-y)^2}}\right)^2 \\&- \frac{1}{8\pi}(\lambda'(x))\left({\int_{\mathbb{R}}\text{d}y\, \frac{\lambda(y) }{(x-y)^2}}\right)\text{Im} \braket{X^2},
} where $\braket{X^2} \equiv \bra{\mathrm{A}_0}X^2\ket{\mathrm{A}_0}$.
Notice that the second term causes the energy density to be non-zero outside the support of A's smearing. We can see that this is not the case for $d=2$ where $\braket{X^2}=1$ and only the first term of the above survives. However, for $d>2$ the second term gives a non-local contribution to the energy density. This is, A's action puts energy outside of her instantaneous support. While this may seem not reasonable from the point of view of the causality of the interaction model \cite{EMM2015}, we note that the energy that A deposits in the field non-locally decays with the fourth power of the distance to A's support. This means that, in the large teleportation distance limit, the interaction model becomes approximately causal with as much accuracy as desired. Most importantly, the teleported energy \eqref{eq:delta_E_spatial} decays slower than this non-locality.

Indeed, comparing the second term of \eqref{eq:E_dens_A} to the second term of \eqref{eq:delta_E_spatial}, we notice that the non-locality of \eqref{eq:E_dens_A} scales as $\mathcal{O}(L^{-4})$ (where $L$ is the spatial distance to the support of $\lambda$, the spatial smearing of A's coupling) while the teleportation term of \eqref{eq:delta_E_spatial} scales as $\mathcal{O}(L^{-2})$. This means that for $L$ sufficiently large, we have $\mathcal{O}(L^{-2})\gg\mathcal{O}(L^{-4})$, i.e., our teleported energy outweighs the energy leaked by the non-locality of A's action on the field. Therefore, we see that it is in principle possible to recover a proper QET scheme in the asymptotic limit.
To have a genuine energy teleportation scheme in this asymptotic limit, we should be able to compensate for the distance decay of B's ability to teleport energy with an increase of A's coupling strength, all while keeping the non-local term of the interaction much smaller. From equations \eqref{eq:delta_E_spatial} and \eqref{eq:E_dens_A} (and appendix \ref{ApD})  these two conditions translate into
\begin{equation}\label{eq:3_ineq}
\frac{\mu_0}{\sigma_B}\ll\frac{\lambda_0}{L^2}\ll1,
\end{equation}
where $\sigma_B$ is B's detector smearing support lengthscale (the size of his detector). Note that we assume $d$ is sufficiently large so that $|\text{Im}(\Gamma)|\approx 1$ (discussed in \ref{sec:dim_vs_E}) in the above.

These two conditions guarantee that we can do QET (the second summand of \eqref{eq:delta_E_spatial} is larger in magnitude than the first) and that the QET scheme is not due to non-locality of the interaction Hamiltonian. We can now see whether we can scale our setup so that unbounded QET is possible, if at least in the asymptotic limit.  For this, we need the second summand in \eqref{eq:delta_E_spatial} to be able to scale unboundedly as we scale the rest of the parameters, being careful that at the same time the size of B's detector, $\sigma_B$, scales slower than the separation between detectors.

For illustration, let us rescale the problem parameters so that all of them are expressed in terms of the dimensionless rescaling parameter $\eta$ so that $L\sim \eta$. If now $\lambda_0\sim \eta^{2-\epsilon}$, $\mu_0\sim \eta^{2\epsilon}$, $\sigma_B\sim \eta^{4\epsilon}$ both inequalities can be expressed as
\begin{equation}\eta^{-\epsilon}\ll1
\end{equation}
which is true for any $\epsilon>0$ for large enough $\eta$. Now, for this particular choice of scaling the teleported energy scales as 
\begin{equation}\frac{\lambda_0\mu_0}{L^2}\sim \eta^{\epsilon}\end{equation}
which means it can be made arbitrarily large increasing $\eta$. At the same time if we choose $0<\epsilon<1/4$ we get that 
\begin{equation}\sigma_B\sim \eta^{4\epsilon}<\eta\end{equation}
for large enough $\eta$, so, in other words, we can scale up the separation between the detectors by a factor $\eta$ if we scale up at the same time the coupling strengths of A and B by a factor $\eta^{2-\epsilon}$ and $\eta^{2\epsilon}$ respectively while at the same time  scaling the width of B's detector $\sigma_B$ as $\eta^{4\epsilon}$. For the separation between the detectors to scale up faster than the length of detector B we need to demand that $\epsilon<1/4$. As a result, the teleported energy scales as $\eta^\epsilon$ which in the asymptotic limit where locality is recovered exactly, $\eta\rightarrow \infty$, is unbounded.

Note that in the above analysis, we would also need to scale up the dimensions $d$ of the probe at least as fast as the scaling of $\lambda_0$ to get a value of $\Gamma \rightarrow 1$, as was discussed in section \ref{sec:dim_vs_E}. To be sure that this factor does not affect the above analysis, we require the scaling $d\sim \eta^2>\eta^{2-\epsilon}\sim \lambda_0$, hence for sufficiently large $\eta$, we have $|\text{Im}(\Gamma)| \approx 1$. 

Notice that, in this asymptotic limit, one can teleport an arbitrarily large amount energy from A to B. Since the initial field state was the vacuum, this implies that, asymptotically, B deposits an arbitrary large amount of negative energy in the field in and near his spatial support. However, the \textit{density} of this negative energy does in fact scale down as we scale up the teleported energy, namely 
\begin{equation}
\frac{\lambda_0\mu_0}{L^2 \sigma_B}\sim \eta^{-3\epsilon}.
\end{equation}
Indeed, if the initial field state is the vacuum, in the asymptotic regime where  the QET protocol becomes local (i.e., where the non-localities of the interaction are sufficiently suppressed), it is still not possible to deposit an arbitrarily negative energy density in the field, as expected from Flanagan's quantum inequality \cite{Flanagan97}. This shows that teleporting arbitrarily large amounts of energy through QET is compatible with Flanagan's theorem in a suitable asymptotic regime.

For completeness, we can calculate the energy density after B's interaction. B's state after his interaction is given by,
\begin{equation}
\ket{\Psi_3} = \sum_{ij} \braket{z_i|x_j}\!\braket{x_j|\mathrm{A}_0} \ket{z_i} D(\bm{\xi}_i)\ket{\bm{\Lambda_j}}
\end{equation}
and we can calculate the energy density in the field for this state, with some work we get
\spliteq{\label{eq:B_E_dens}
\!\!\!\!\!\!E_B(x) &\equiv \bra{\Psi_3} I \otimes \varepsilon_+(x)\ket{\Psi_3} = \overbrace{\left|\mathcal{I}_A\right|^2- \text{Re}\left[\Gamma_{20}\, \mathcal{I}_A^2\right]}^{E_A(x+T)}\\ &+ \left|\mathcal{I}_B\right|^2- \text{Re}\left[\Gamma_{02}\, \mathcal{I}_B^2\right]-2 \text{Re}\left[\Gamma_{11}\, \mathcal{I}_A\mathcal{I}_B\right]\\& - 2 \text{Re}\left[\Gamma_{1(-1)}\, \mathcal{I}_A\mathcal{I}^*_B\right]
}
where 
\begin{align}
\mathcal{I}_A &\equiv \tfrac{1}{\sqrt{4\pi}}\int_{\mathbb{R}^+}\!\!\!\! \text{d}\omega \sqrt{\omega} \alpha_\omega^* e^{\ii \omega (x+T)} \\
\mathcal{I}_B &\equiv \tfrac{1}{\sqrt{4\pi}}\int_{\mathbb{R}^+}\!\!\!\!\text{d}\omega \sqrt{\omega} \beta_\omega^* e^{\ii \omega x}
\end{align}
\vspace{-4.2mm}
\begin{equation}
\Gamma_{mn}\equiv e^{\left(e^{-2\pi \ii n/d}-1\right)\lVert\bm{\alpha}\rVert^2}\bra{\mathrm{A}_0}X^mZ^n\ket{\mathrm{A}_0},
\end{equation}
where $m,n \in\mathbb{Z}_d$ (recall that the Weyl operators are unitary; $X^{-1} =X^\dagger$, $Z^{-1} =Z^\dagger$).
the spatial representation our integrals are
\begin{align}
\mathcal{I}_A &= -\frac{i}{4}\lambda'(x+T) -\frac{1}{4\pi}\int_{\mathbb{R}}\! \text{d}y \frac{\lambda(y)}{(x-y+T)^2}\\
\mathcal{I}_B &= -\frac{i}{4}\mu(x) +\frac{1}{4\pi}\int_{\mathbb{R}}\! \text{d}y \frac{\mu(y)}{(x-y)}.
\end{align}
The first two terms of \eqref{eq:B_E_dens} give us A's energy density, translated to the left by $T$. The third and fourth terms are B's investment in the field from switching his detector, similar to A's. The fifth and sixth terms are the interesting ones since they can create negative energy densities. The sixth term is our teleportation term from our usual calculation, before being spatially integrated. The fifth term is similar to our teleportation term. In the appropriate scaling limit (discussed above), the last two terms will dominate in the vicinity of B's support, creating a locally negative energy density in the field.

\subsection{Information Theoretic Considerations} 

\subsubsection{Information Redundancy}
We have shown that by scaling up the amount of information exchanged between A and B along with the amount of energy injected in the field it is possible to obtain arbitrarily large teleported energies. 

One might wonder what would happen if we leave A's coupling strength fixed (and the smearing) while scaling up the dimensions, i.e. keep $\lambda(x)$ fixed while increasing $d$. Our bound for the teleportation term's coefficient \eqref{eq:Gamma_bound} becomes
\begin{equation}
\!\!\!\!\left|\text{Im}(\Gamma)\right|  \geq \cos(\tfrac{\pi}{d})e^{\left(\cos\left(\tfrac{2\pi}{d}\right)-1\right)\lVert\bm{\alpha}\rVert^2} \overset{d\rightarrow\infty}{\longrightarrow} 1.
\end{equation}
We see that simply increasing the dimensions has diminishing returns. As $d\gg \lVert\bm{\alpha}\rVert$, adding more dimensions to our probe does not yield much of an advantage in terms of energy extracted. In other words, increasing the number of potential outcomes of our measurement does not necessarily mean that it will yield more information about the field state that is useful for QET.

Although an increase in the dimensions of the probe without a corresponding increase in coupling strength has a diminishing advantage in terms of energy teleported, it can still be useful to consider higher dimensions for a given fixed coupling strength to counter the effects of noise in the communication channel.
Say that the quantum channel between A and B were to have quantum noise in the form of Weyl operators, i.e., with a probability $p(a,b)$, $Z^aX^b$ is applied onto the qudit in (\ref{eq:after_transfer}), then on average our teleportation term in (\ref{eq:Delta_E}) would pick up a factor of $\sum_{a,b\in\mathbb{Z}_d}p(a,b)\cos(2\pi b/d)$. If most errors have $b\ll d$, then we are still capable of fairly efficient QET. The analogous statement for the classical communication scenario would be that the new QET protocol can withstand noise as long as it only causes small shifts relative to the size of the alphabet. This fault tolerance further highlights that increasing the dimensions of the probe without scaling other parameters causes our measurement information to increase its 'redundancy' (in the sense of error correction). 

\subsubsection{Entanglement Harvesting}

A possible explanation for this redundancy would be that since A's energy injected into the field is related to the amount of entanglement acquired between her probe and the field by her measurement, keeping that energy fixed imposes a bound to the amount of probe-field entanglement that can be achieved by the probe-field interaction. Increasing the dimensions of the probe would then simply allow for the approach to that bound. Below we provide some mathematical and geometrical intuition to support the above explanation.

Looking at equation \eqref{eq:A_state}, we see that the probe and the field are nearly in a Schmidt decomposition; if the $\{\ket{\bm{\zeta}_j} = \ket{e^{2\pi\ii j/d}\bm{\alpha}}: j\in\mathbb{Z}_d\}$ were orthogonal, then the probe would be maximally entangled with the field. Note that these coherent states are the ones superposed in figure \ref{fig:catstates}. The inner product between any two of these states,
$\braket{\bm{\zeta}_m|\bm{\zeta}_k}$ is the same as in equation \eqref{eq:inner_prod}. The smaller these inner products are, the closer to maximally entangled the probe is. These inner products increase when increasing $d$ or decreasing $\lVert\bm{\alpha}\rVert^2$. This supports our intuition that a more entanglement requires greater energy injected and that increasing the dimensions while keeping the energy fixed only worsens the sub-maximality of the entanglement. 

Geometrically, looking at figure \ref{fig:catstates}, since each coherent state has a fixed variance in phase space (saturating Heisenberg uncertainty), then a limited number of coherent states can be resolved for a fixed radius when these are placed at the vertices of a regular $d$-gon (of said radius). As manifested in \eqref{eq:A_state}, the probe's state is perfectly correlated with each coherent state, hence these need to be well discernible in order for measurements of the probe to transitively give us accurate knowledge of the field's phase. To fit more minimally overlapping coherent states, the radius ($|\alpha|$, analogous to $\lVert\bm{\alpha}\rVert$) needs to be increased. If we were to increase the radius, we would intuitively expect the maximum number of minimally overlapping coherent states to scale as the perimeter of a circle, hence linearly in the radius. We thus have geometrically justified our linear scaling result in \eqref{eq:lrg_theta}.

\subsection{Energy Conditions} 

Flanagan's theorem \cite{Flanagan97} sets limits to the amount of energy that can be teleported by a local QET protocol. In our protocol we overcome the limitation imposed by the theorem by allowing for a non-local interaction between the field and A and B's probes. In the asymptotic limit where A and B are far apart and the support of B is scaled up, the non-locality contribution to the energy extracted by B becomes arbitrarily small as compared to the teleported energy.

As discussed above, we calculated that the field admits a locally negative energy density near B, i.e., averaged over B's locale, $\text{supp}(\mu)$, we have $\braket{:T_{++}:}<0$. Although this is locally a violation of the pointwise Null Energy Condition (NEC) \cite{Hawking1973}, which is not uncommon for quantum fields, the 
Averaged Null Energy Condition (ANEC) \cite{Kelly2014} (integrated over a complete light sheet) is respected. This is because the energy recovered by B is less than or equal to A's measurement energy investment, $\Delta E\leq E_A$. In terms of bounds related to the spatial profile, QET obeys Ford's quantum inequalities \cite{Ford1994}.
A bound potentially relevant to our  new protocol is the Quantum Null Energy Condition \cite{Bousso2015}, since it provides a pointwise lower bound for the stress energy tensor which is related to the entanglement entropy of the field. 

It would be interesting to verify if this bound can be saturated via the new qudit-QET approach.

\section{Conclusions}

We have shown that two distant agents A and B equipped with qudit probes which interact with a massless scalar field on a compactly supported region of space, can teleport an arbitrary amount of energy from one to another in a suitable asymptotic regime, without any energy flowing from A to B, and by means of only localized operations on the field and classical or quantum communication. The scaling up of the amount of teleported energy is achieved by scaling protocol parameters, and in particular by scaling up the amount of information about the field obtained by A and sent to B.

Remarkably, in order for A to be able to harvest increasingly more information from the field that is useful to teleport an arbitrarily large amount of energy, it is not sufficient to simply increase the dimension of the usual Unruh-DeWitt detector keeping the interaction local, as this would contradict Flanagan's theorem. Instead, we overcome this limitation by introducing a weak non-locality in the interaction Hamiltonian. The non-locality is such that a) it allows us to circumvent the constraints of Flanagan's quantum inequality, and b) it is such it becomes negligible in a suitable asymptotic regime.

Additionally, we have shown that the quantum and classical communication versions of the protocols yield the same outcome in terms of energy extracted on average. This further corroborates the findings of \cite{Nambu2010} that the key resource for QET are the classical correlations of the field vacuum.

Our results are of interest not only in this asymptotic regime. There are setups where non-localities in the interaction Hamiltonian appear naturally. An obvious example is solid-state physics where the speed of sound can be much smaller than the speed of light, which implies that there can be interactions with effective non-localities similar to \eqref{eq:A_int_dens} and \eqref{eq:B_int_dens}. Additionally, Hamiltonians with exactly the same degree of non-locality appear in quantum-optical setups whenever the rotating-wave approximation is used \cite{EMM2015}. Our results may therefore be useful in the context of quantum information processing. Another scenario where non-locality is expected to appear is Planck-scale physics.

For experimental realizations, our modified protocol for QET requires qudit systems. Qudits have various experimental realizations, e.g., in quantum optics, often involving the orbital angular momentum states of photons \cite{Molina-Terriza2007, OSullivan-Hale2005, Neves2005}.  There have also been theoretical proposals for quantum error corrected realizations of stable qudits and their logical Weyl operators, either using a stabilizer subspace of harmonic oscillator states \cite{Gottesman2000}, or using a topological quantum memory employing abelian anyons \cite{Wootton2011}.

\textit{Acknowledgements.-} The authors are greatly thankful to Masahiro Hotta for his always insightful and helpful remarks. We also thank Jos\'e Trevis\'on for his valuable feedback. GVA thanks Michael Broughton for technical assistance with Python. E.M-M and AK acknowledge funding from the NSERC Discovery program. The plot in this paper was generated using  QuTiP 2 \cite{Johansson2013}.

\appendix
\begin{widetext}
\section{Energy Calculations}\label{ApA}

We start with the calculation of A's investment of energy into the field from her interaction, i.e. a proof of equation \eqref{eq:E_A}. 
\spliteq{
E_A &=\bra{\Psi_2}I\otimes H\ket{\Psi_2} = \sum_{m=0}^{d-1}\sum_{k=0}^{d-1} (\braket{\textrm{A}_0|x_m}\underbrace{\braket{x_m|x_k}}_{\delta_{mk}}\braket{x_k|\textrm{A}_0} ) \bra{\bm{\Lambda}_m}H\ket{\bm{\Lambda}_k}\\
&= \sum_{k=0}^{d-1} (\braket{\textrm{A}_0|x_k}\braket{x_k|\textrm{A}_0} ) \bra{\bm{\Lambda}_k}\int_0^\infty \text{d}\omega\ \omega a^\dagger_\omega a_\omega\ket{\bm{\Lambda}_k}\\
&= \int_0^\infty \text{d}\omega\ \omega \sum_{k=0}^{d-1} (\braket{\textrm{A}_0|x_k}\braket{x_k|\textrm{A}_0} ) \underbrace{|\Lambda_{k\omega}|^2}_{=|\alpha_{\omega}|^2}\cancelto{1}{\braket{\bm{\Lambda}_k|\bm{\Lambda}_k}}\\
&= \int_0^\infty \text{d}\omega\ \omega  \bra{\textrm{A}_0}\cancelto{I}{\left(\sum_{k=0}^{d-1}\ket{x_k}\braket{x_k}\right)}\ket{\textrm{A}_0} |\alpha_{\omega}|^2\\
&= \int_0^\infty \text{d}\omega\ \omega  |\alpha_{\omega}|^2 = \bra{\bm\alpha}H\ket{\bm\alpha}\\
}
By conservation of energy by free evolution of the field, we know that the energy right before B's interaction is still given by $E_A$. We can compute the total energy in the field right after B's interaction, 

\spliteq{
&E_A + \Delta E \equiv \bra{\Psi_3}I\otimes H\ket{\Psi_3}=\sum_{m=0}^{d-1}\sum_{b=0}^{d-1}  \sum_{k=0}^{d-1} \braket{\textrm{A}_0|x_m}\braket{x_m|z_b}\braket{z_b|x_k}\braket{x_k|\textrm{A}_0} \\ &\qquad\qquad\qquad\qquad\qquad\qquad\qquad\qquad\qquad\qquad\times \bra{\bm{\Lambda}_m}D^\dagger(\bm{\xi}_b)H D(\bm{\xi}_b)\ket{\bm{\Lambda}_k}
\\&=\sum_{m,b,k=0}^{d-1} \braket{\textrm{A}_0|x_m}\braket{x_m|z_b}\braket{z_b|x_k}\braket{x_k|\textrm{A}_0} \bra{\bm{\Lambda}_m}D^\dagger(\bm{\xi}_b)\int_0^\infty \text{d}\omega\ \omega a^\dagger_\omega a_\omega D(\bm{\xi}_b)\ket{\bm{\Lambda}_k}
\\&=\int_0^\infty \text{d}\omega\ \omega\sum_{m,b,k=0}^{d-1} \braket{\textrm{A}_0|x_m}\braket{x_m|z_b}\braket{z_b|x_k}\braket{x_k|\textrm{A}_0}\bra{\bm{\Lambda}_m} (a^\dagger_\omega +\xi^*_{b\omega})(a_\omega+ \xi_{b\omega})\ket{\bm{\Lambda}_k}
\\&=\int_0^\infty \text{d}\omega\ \omega\sum_{m,b,k=0}^{d-1} \braket{\textrm{A}_0|x_m}\braket{x_m|z_b}\braket{z_b|x_k}\braket{x_k|\textrm{A}_0} \braket{\bm{\Lambda}_m|\bm{\Lambda}_k}(\Lambda^*_{m\omega} +\xi^*_{b\omega})(\Lambda_{k\omega}+ \xi_{b\omega})
\\&=\int_0^\infty \text{d}\omega\ \omega\sum_{m,b,k=0}^{d-1} \braket{\textrm{A}_0|x_m}\braket{x_m|z_b}\braket{z_b|x_k}\braket{x_k|\textrm{A}_0} \braket{\bm{\Lambda}_m|\bm{\Lambda}_k}(\underbrace{\Lambda^*_{m\omega} \Lambda_{k\omega}+ \xi^*_{b\omega}\xi_{b\omega}}_{\textrm{I}}+\underbrace{ \Lambda^*_{m\omega}\xi_{b\omega}+\xi^*_{b\omega}\Lambda_{k\omega}}_{\textrm{II}})
}
To help ourselves in our calculation, we divide the different terms in order to better conquer them. We have the ``symmetric'' and the ``cross'' terms (\textrm{I} and \textrm{II} respectively). Keep in mind then that $E_A + \Delta E =\textrm{I} +\textrm{II}$. Let us start with the symmetric ones:
\spliteq{
\textrm{I} &= \int_0^\infty \text{d}\omega\ \omega\sum_{m,b,k=0}^{d-1} \braket{\textrm{A}_0|x_m}\braket{x_m|z_b}\braket{z_b|x_k}\braket{x_k|\textrm{A}_0} \braket{\bm{\Lambda}_m|\bm{\Lambda}_k}(\Lambda^*_{m\omega} \Lambda_{k\omega}+ \xi^*_{b\omega}\xi_{b\omega})\\
 &= \int_0^\infty \text{d}\omega\ \omega\sum_{m,k=0}^{d-1} \braket{\textrm{A}_0|x_m}\underbrace{\bra{x_m}\cancelto{I}{\left(\sum_{b=0}^{d-1}\ket{b}\bra{b}\right)}\ket{x_k}}_{\delta_{mk}}\braket{x_k|\textrm{A}_0} \braket{\bm{\Lambda}_m|\bm{\Lambda}_k}(\Upsilon^{m-k}|\alpha_\omega|^2+ |\beta_\omega|^2)\\
  &= \int_0^\infty \text{d}\omega\ \omega \cancelto{1}{\left(\bra{\textrm{A}_0}{\left(\textstyle{\sum_{k=0}^{d-1}}\ket{x_k}\bra{x_k}\right)}\ket{\textrm{A}_0}\right)} (|\alpha_\omega|^2+ |\beta_\omega|^2)\cancelto{1}{\braket{\bm{\Lambda}_k|\bm{\Lambda}_k}}\\
  &= \int_0^\infty \text{d}\omega\ \omega \ (|\alpha_\omega|^2+ |\beta_\omega|^2)= \underbrace{\bra{\bm\alpha}H\ket{\bm\alpha}}_{=E_A}+\bra{\bm\beta}H\ket{\bm\beta} = E_A+\bra{\bm\beta}H\ket{\bm\beta}.
}
We see that these terms yield the independent energy of the coherent states generated by A and B's interactions. This would be the energy in the field if A and B would both do their interactions without being causally related. The interesting behaviour of QET hence lies in the cross terms. Indeed, the work extracted (the negative energy) will be in the \textrm{II} terms,
\begin{equation}
\Delta E = E_A + \Delta E -E_A = \textrm{I} +\textrm{II} -E_A = \underbrace{\bra{\bm\beta}H\ket{\bm\beta}}_{\geq 0}+\underbrace{\textrm{II}}_{\overset{?}{<}0}.
\end{equation}
\\

Before we attack the rest of the calculation, note the following identity which we be put to use: In general, for operators $A,B$ and scalar $\gamma$,

\begin{equation}
AB = \gamma BA \implies  AB^n = \gamma^n B^n A \implies Ae^B = e^{\gamma B}A.
\end{equation}
Finally, before we proceed, recall our definition
\begin{equation}
\lVert\bm{\alpha}\rVert^2 \equiv \int_0^\infty \text{d}\omega\  |\alpha_\omega|^2.
\end{equation}
We now compute the cross terms,
\spliteq{
\textrm{II} &= \int_0^\infty \text{d}\omega\ \omega\sum_{m,b,k=0}^{d-1} \braket{\textrm{A}_0|x_m}\braket{x_m|z_b}\braket{z_b|x_k}\braket{x_k|\textrm{A}_0} \underbrace{\braket{\bm{\Lambda}_m|\bm{\Lambda}_k}}_{\exp\left((\Upsilon^{m-k}-1)\lVert\bm{\alpha}\rVert^2\right)}( \Lambda^*_{m\omega}\xi_{b\omega}+\xi^*_{b\omega}\Lambda_{k\omega})\\
&=  \int_0^\infty \text{d}\omega\ \omega\sum_{m,b,k=0}^{d-1} \braket{\textrm{A}_0|x_m}\braket{x_m|z_b}\braket{z_b|x_k}\braket{x_k|\textrm{A}_0} e^{(\Upsilon^{m-k}-1)\lVert\bm{\alpha}\rVert^2}(\underbrace{\Upsilon^{m-b} \alpha^*_{\omega}(T)\beta_{\omega}}_{(\textrm{i})}+\underbrace{\Upsilon^{b-k} \alpha_{\omega}(T)\beta^*_{\omega}}_{\textrm{(ii)}})\\
}
starting with the first one, 
\spliteq{
\textrm{(i)}&= \int_0^\infty \text{d}\omega\ \omega\sum_{m,b,k=0}^{d-1} \braket{\textrm{A}_0|x_m}\braket{x_m|z_b}\braket{z_b|x_k}\braket{x_k|\textrm{A}_0} e^{(\Upsilon^{m-k}-1)\lVert\bm{\alpha}\rVert^2}\Upsilon^{m-b}\alpha^*_{\omega}(T)\beta_{\omega} \\
&= e^{-\lVert\bm{\alpha}\rVert^2}\left(\int_0^\infty \text{d}\omega\ \omega \alpha^*_{\omega}(T)\beta_{\omega}\right)\\
&\qquad\times\bra{\textrm{A}_0}\left[\sum_{m=0}^{d-1}\Upsilon^{m}\ket{x_m}\bra{x_m}\underbrace{\left(\sum_{b=0}^{d-1}\Upsilon^{-b}\ket{b}\bra{b}\right)}_{Z^\dagger}\underbrace{\left(\sum_{k=0}^{d-1}e^{(\Upsilon^{m}\lVert\bm{\alpha}\rVert^2)\Upsilon^{-k}}\ket{x_k}\bra{x_k}\right)}_{\exp\left(\Upsilon^m \lVert\alpha\rVert^2 X^\dagger\right)}\right]\ket{\textrm{A}_0}\\
&= e^{-\lVert\bm{\alpha}\rVert^2}\left(\int_0^\infty \text{d}\omega\ \omega \alpha^*_{\omega}(T)\beta_{\omega}\right)\bra{\textrm{A}_0}\left[\sum_{m=0}^{d-1}\Upsilon^{m}\ket{x_m}\bra{x_m}Z^\dagger e^{\Upsilon^m \lVert\alpha\rVert^2 X^\dagger}\right]\ket{\textrm{A}_0}\\
&= e^{-\lVert\bm{\alpha}\rVert^2}\left(\int_0^\infty \text{d}\omega\ \omega \alpha^*_{\omega}(T)\beta_{\omega}\right)\bra{\textrm{A}_0}\left[\sum_{m=0}^{d-1}\Upsilon^{m}\ket{x_m}\bra{x_m}e^{\Upsilon^{m+1} \lVert\alpha\rVert^2 X^\dagger}Z^\dagger\right]\ket{\textrm{A}_0}\\
&= e^{-\lVert\bm{\alpha}\rVert^2}\left(\int_0^\infty \text{d}\omega\ \omega \alpha^*_{\omega}(T)\beta_{\omega}\right)\bra{\textrm{A}_0}\left[\sum_{m=0}^{d-1}\Upsilon^{m}\ket{x_m}\bra{x_m}e^{\Upsilon^{\cancel{m}+1} \lVert\alpha\rVert^2 \cancel{\Upsilon^{-m}}}Z^\dagger\right]\ket{\textrm{A}_0}\\
&= e^{(\Upsilon-1)\lVert\bm{\alpha}\rVert^2}\left(\int_0^\infty \text{d}\omega\ \omega \alpha^*_{\omega}(T)\beta_{\omega}\right)\bra{\textrm{A}_0}\underbrace{\left(\sum_{m=0}^{d-1}\Upsilon^{m}\ket{x_m}\bra{x_m}\right)}_{X} Z^\dagger\ket{\textrm{A}_0}\\
&=e^{(\Upsilon-1)\lVert\bm{\alpha}\rVert^2}\left(\int_0^\infty \text{d}\omega\ \omega e^{i\omega T}\alpha^*_{\omega}\beta_{\omega}\right)\bra{\textrm{A}_0}X Z^\dagger\ket{\textrm{A}_0}
}
where in the above we use the fact that $Z^\dagger X^\dagger = \Upsilon X^\dagger Z^\dagger$. The second cross term is now easy to compute since it is the complex conjugate of the first,
\spliteq{
\textrm{(ii)}&= e^{(\Upsilon^{-1}-1)\lVert\bm{\alpha}\rVert^2}\left(\int_0^\infty \text{d}\omega\ \omega e^{-i\omega T}\alpha_{\omega}(T)\beta^*_{\omega}\right)\bra{\textrm{A}_0}ZX^\dagger \ket{\textrm{A}_0}.
}
Thus the cross-terms end up summing up to the following, 
\spliteq{
\textrm{II} = \textrm{(i)}+\textrm{(ii)}= 2\text{Re}[\textrm{(i)}] = 2\text{Re}\Bigg[\left(\int_0^\infty \text{d}\omega\ \omega e^{i\omega T}\alpha^*_{\omega}\beta_{\omega}\right)\bra{\textrm{A}_0}X Z^\dagger\ket{\textrm{A}_0}e^{(\Upsilon-1)\lVert\bm{\alpha}\rVert^2}\Bigg].
}
Putting all of the above together we recover our results from equation \eqref{eq:Delta_E}.

\section{Stress-Energy Tensor Calculation}\label{ApB}

The left-moving energy density of the field is given by 

\spliteq{
\varepsilon_+(x^+) &\equiv  \frac{1}{2}: \Pi_+(x^+)^2: = \frac{1}{2}:\int\limits_{(\mathbb{R}^+)^2}\text{d}\omega \text{d}\omega' \frac{ \sqrt{\omega\omega'}}{4\pi} \left[a^{\dagger}_\omega e^{\ii\omega x^+}-a_\omega e^{-\ii\omega x^+}\right]\left[a_{\omega'} e^{-\ii\omega' x^+}-a^{\dagger}_{\omega'} e^{\ii\omega' x^+}\right] :\\
&= \frac{1}{2}\int\limits_{(\mathbb{R}^+)^2}\text{d}\omega \text{d}\omega' \frac{ \sqrt{\omega\omega'}}{4\pi} :\left[a^{\dagger}_\omega a_{\omega'} e^{\ii(\omega-\omega') x^+}-a^{\dagger}_\omega a^{\dagger}_{\omega'} e^{\ii(\omega+\omega') x^+}-a_\omega a_{\omega'} e^{-\ii(\omega+\omega') x^+}+a_\omega a^{\dagger}_{\omega'} e^{-\ii(\omega-\omega') x^+}\right] :\\
&=\int\limits_{(\mathbb{R}^+)^2}\text{d}\omega \text{d}\omega' \frac{ \sqrt{\omega\omega'}}{8\pi} \left[a^{\dagger}_\omega a_{\omega'} e^{\ii(\omega-\omega') x^+}+ a^{\dagger}_{\omega'}a_\omega e^{-\ii(\omega-\omega') x^+}-a^{\dagger}_\omega a^{\dagger}_{\omega'} e^{\ii(\omega+\omega') x^+}-a_\omega a_{\omega'} e^{-\ii(\omega+\omega') x^+}\right]\\
&=\int\limits_{(\mathbb{R}^+)^2}\text{d}\omega \text{d}\omega' \frac{ \sqrt{\omega\omega'}}{8\pi} \left[2a^{\dagger}_\omega a_{\omega'} e^{\ii(\omega-\omega') x^+}-a^{\dagger}_\omega a^{\dagger}_{\omega'} e^{\ii(\omega+\omega') x^+}-a_\omega a_{\omega'} e^{-\ii(\omega+\omega') x^+}\right].
}
For a fixed time, say $t=0$, we can integrate the above over all space to obtain the total Hamiltonian (for left-movers):
\spliteq{
H_+ &\equiv \int_{\mathbb{R}}\!\! \text{d}x\, \varepsilon_+(x) = \int_{(\mathbb{R}^+)^2}\!\!\!\!\!\!\!\!\text{d}\omega \text{d}\omega' \frac{ \sqrt{\omega\omega'}}{8\pi} \left[4\pi a^{\dagger}_\omega a_{\omega'} \delta(\omega-\omega')-\cancel{2\pi a^{\dagger}_\omega a^{\dagger}_{\omega'} \delta(\omega+\omega')}-\cancel{2\pi a_\omega a_{\omega'} \delta(\omega+\omega')}\right]\\
&= \int_{(\mathbb{R}^+)^2}\!\!\!\!\!\!\!\!\text{d}\omega\, \frac{\omega}{2}a^{\dagger}_\omega a_{\omega},
}
we see only the first term survives. This is a standard result which is worth reiterating since it shows that our energy density calculations and total energy calculations should be in agreement as long as we use the above.

 The state right after A's interaction (at $t=0^+$) is given by $\ket{\Psi_1}$. We can compute the value of the energy density immediately after this interaction (note $x^+ =x$ since $t=0$).
\spliteq{
&\!E_A(x) =\bra{\Psi_1}I\otimes \varepsilon_+(x) \ket{\Psi_1} = \sum_{j=0}^{d-1}\sum_{i=0}^{d-1} (\braket{\mathrm{A}_0|x_j}\underbrace{\braket{x_j|x_i}}_{\delta_{ji}}\braket{x_i|\mathrm{A}_0} ) \bra{\bm{\zeta}_j}\varepsilon_+(x) \ket{\bm{\zeta}_i}\\
&= \sum_{j=0}^{d-1} (\braket{\mathrm{A}_0|x_j}\!\braket{x_j|\mathrm{A}_0} ) \bra{\bm{\zeta}_j}\int\limits_{(\mathbb{R}^+)^2}\text{d}\omega \text{d}\omega' \frac{ \sqrt{\omega\omega'}}{8\pi} \Big[2a^{\dagger}_\omega a_{\omega'} e^{\ii(\omega-\omega') x^+}-a^{\dagger}_\omega a^{\dagger}_{\omega'} e^{\ii(\omega+\omega') x}-a_\omega a_{\omega'} e^{-\ii(\omega+\omega') x}\Big]\ket{\bm{\zeta}_j}\\
&= \int\limits_{(\mathbb{R}^+)^2}\text{d}\omega \text{d}\omega' \frac{ \sqrt{\omega\omega'}}{8\pi}\sum_{j=0}^{d-1} (\braket{\mathrm{A}_0|x_j}\!\braket{x_j|\mathrm{A}_0} )\cancel{\braket{\bm{\zeta}_j|\bm{\zeta}_j}}  \Big[2\zeta^*_{j\omega} \zeta_{j\omega'}e^{\ii(\omega-\omega') x}-2\text{Re}\left(\zeta_{j\omega} \zeta_{j\omega'} e^{-\ii(\omega+\omega') x}\right)\Big]\\
&= \int\limits_{(\mathbb{R}^+)^2}\text{d}\omega \text{d}\omega' \frac{ \sqrt{\omega\omega'}}{4\pi}\sum_{j=0}^{d-1} \braket{\mathrm{A}_0|x_j}\!\braket{x_j|\mathrm{A}_0}  \Big[\alpha^*_{\omega} \alpha_{\omega'}e^{\ii(\omega-\omega') x}-\text{Re}\left(e^{-4\pi\ii j/d}\alpha_{\omega} \alpha_{\omega'} e^{-\ii(\omega+\omega') x}\right)\Big]\\
&=\int\limits_{(\mathbb{R}^+)^2}\text{d}\omega \text{d}\omega' \frac{ \sqrt{\omega\omega'}}{4\pi}  \left[\alpha^*_{\omega} \alpha_{\omega'}e^{\ii(\omega-\omega') x}-\text{Re}\left(\textstyle\sum_{j}e^{-4\pi\ii j/d} \braket{\mathrm{A}_0|x_j}\braket{x_j|\mathrm{A}_0}\alpha_{\omega} \alpha_{\omega'} e^{-\ii(\omega+\omega') x}\right)\right]\\
&=\int\limits_{(\mathbb{R}^+)^2}\text{d}\omega \text{d}\omega' \frac{ \sqrt{\omega\omega'}}{4\pi}  \left[\alpha^*_{\omega} \alpha_{\omega'}e^{\ii(\omega-\omega') x}-\text{Re}\left( \bra{\mathrm{A}_0}X^{\dagger 2}\ket{\mathrm{A}_0}\alpha_{\omega} \alpha_{\omega'} e^{-\ii(\omega+\omega') x}\right)\right]\\
&=\int\limits_{(\mathbb{R}^+)^2}\text{d}\omega \text{d}\omega' \frac{ \sqrt{\omega\omega'}}{4\pi}  \alpha^*_{\omega} \alpha_{\omega'}e^{\ii(\omega-\omega') x}-\int\limits_{(\mathbb{R}^+)^2}\text{d}\omega \text{d}\omega' \frac{ \sqrt{\omega\omega'}}{4\pi} \text{Re}\left( \bra{\mathrm{A}_0}X^{\dagger 2}\ket{\mathrm{A}_0}\alpha_{\omega} \alpha_{\omega'} e^{-\ii(\omega+\omega') x}\right)\\
&=\int\limits_{(\mathbb{R}^+)^2}\text{d}\omega \text{d}\omega' \frac{ \omega\omega'}{(4\pi)^2}  \tilde{\lambda}^*(\omega) \tilde{\lambda}(\omega')e^{\ii(\omega-\omega') x}-\int\limits_{(\mathbb{R}^+)^2}\text{d}\omega \text{d}\omega' \frac{ \omega\omega'}{(4\pi)^2} \text{Re}\left( \bra{\mathrm{A}_0}X^{\dagger 2}\ket{\mathrm{A}_0} \tilde{\lambda}(\omega) \tilde{\lambda}(\omega') e^{-\ii(\omega+\omega') x}\right)\\
&=\frac{ 1}{(4\pi)^2}\left|\textstyle\int_{\mathbb{R}^+}\text{d}\omega\,\omega     \tilde{\lambda}(\omega)e^{-\ii\omega x}\right|^2- \frac{ 1}{(4\pi)^2}\text{Re}\Big(\bra{\mathrm{A}_0}X^{\dagger 2}\ket{\mathrm{A}_0}\left(\textstyle\int_{\mathbb{R}^+}\text{d}\omega\,\omega    \tilde{\lambda}(\omega)  e^{-\ii\omega x}\right)^2\Big)\\
&=\frac{ 1}{(4\pi)^2}\left|\textstyle\int_{\mathbb{R}^+}\text{d}\omega\,\omega     \tilde{\lambda}(-\omega)e^{\ii\omega x}\right|^2- \frac{ 1}{(4\pi)^2}\text{Re}\Big(\bra{\mathrm{A}_0}X^{ 2}\ket{\mathrm{A}_0}\left(\textstyle\int_{\mathbb{R}^+}\text{d}\omega\,\omega    \tilde{\lambda}(-\omega)  e^{\ii\omega x}\right)^2\Big)\\
}
The second term seems like it could admit some non-locality for $d>2$, i.e. when $X^{\dagger 2} \neq I$. To see this more clearly, let
\spliteq{
z&\equiv \tfrac{1}{4\pi}\textstyle\int_{\mathbb{R}^+}\text{d}\omega\,\omega     \tilde{\lambda}(-\omega)e^{\ii\omega x} \equiv a+\ii b,\quad\ a,b\in\mathbb{R} \\
\braket{X^2}&\equiv \bra{\mathrm{A}_0}X^{ 2}\ket{\mathrm{A}_0}\equiv g+h\ii,\quad\ g,h\in\mathbb{R}
}
Then, we have
\spliteq{
E_A(x) &= |z|^2 - \text{Re}(\braket{X^2}z^2)\\
&= (1-g)a^2 + (1+g)b^2 + 2abh.
}
Notice that in the $d=2$ case, we have $ \braket{X^2} =1 = g,\ h=0$ and $E_A(x)=2b^2 = 2 [\text{Im}(z)]^2$. As we will see below, the imaginary part of our Fourier integral $z$ is the local one, while the real part is non-local. It seems that for $d>2$ (qudits), we can add energy density to the field non-locally. Before we proceed, note the following integral 
\begin{equation}
\int_{\mathbb{R}^+}\text{d}\omega\,\omega     e^{\ii\omega x} = \int_{\mathbb{R}}\text{d}\omega\,\omega\, \Theta(x)    e^{\ii\omega x}= - \frac{1}{x^2}- \ii \pi \delta'(x)
\end{equation}
we already see that the imaginary part is local (delta function) versus a non-local power-law decay. Now, we can compute our $z$ (convolution and integration by parts),
\spliteq{z = \frac{1}{4\pi}\int_{\mathbb{R}}dy\, \lambda(y) \left(\frac{-1}{(x-y)^2}-\ii \pi \delta'(x-y)\right)= - \frac{1}{4\pi}\int_{\mathbb{R}}dy\, \frac{\lambda(y) }{(x-y)^2} +\ii\frac{\lambda'(x) }{4}  .
}
Thus, as our final expression for the energy density, we have
\begin{equation}
E_A(x) = \tfrac{1}{16\pi^2}(1-\text{Re} \braket{X^2})\left(\textstyle{\int_{\mathbb{R}}dy\, \frac{\lambda(y) }{(x-y)^2}}\right)^2 + \tfrac{1}{16}(1+\text{Re} \braket{X^2})(\lambda'(x))^2 - \tfrac{1}{8\pi}(\lambda'(x))\left(\textstyle{\int_{\mathbb{R}}dy\, \frac{\lambda(y) }{(x-y)^2}}\right)\text{Im} \braket{X^2}.
\end{equation}
Note that the first term is what gives us non-locality, the second term is local term and the third term is also local (since $\lambda'$ transfers it compact support property to the product).
A question arises: can we have a choice of initial state $\ket{\mathrm{A}_0}$ such that we have both efficient QET and remove the non-locality of A? The only way to remove all non-locality is to have $\braket{X^2}=1$, this means we would need to choose $\ket{A_0} = \ket{x_0}$ the eigenstate of $X$ of eigenvalue $1$ when $d>2$. Unfortunately, the teleportation term goes as $\sim\bra{A_0}XZ^\dagger\ket{A_0}$, so if we have $\ket{A_0} = \ket{x_0}$, then 
\begin{equation}
\bra{x_0}XZ^\dagger\ket{x_0} = \braket{x_0|x_{d-1}} = 0
\end{equation}
and we have no teleportation. Note that, our non-locality decays as the distance $L^{-4}$, which is a faster decay than the energy extracted $L^{-2}$.

\section{Hamiltonian non-locality}\label{ApC}
Our interaction Hamiltonian density (omitting the $\lambda(x)$ smearing) is given by 
\spliteq{
\mathcal{H}_\mathcal{A}(x) &\equiv -\ii\ \int_{\mathbb{R}^+} \text{d}\omega \sqrt{\tfrac{\omega}{4\pi}} \left(X a_\omega e^{-\ii\omega x} -X^\dagger  a^\dagger_\omega e^{\ii\omega x}\right).
}
We can decompose the Weyl operators into their Hermitian and anti-Hermitian components, let
\spliteq{
C\equiv \frac{(X+X^\dagger)}{2}, &\quad S\equiv \frac{i(X-X^\dagger)}{2},\\X\equiv \frac{(C-\ii S)}{2}, &\quad X^\dagger\equiv \frac{(C+\ii S)}{2}.
}
Note that when $d=2$, $S=0$.
We can rewrite our interaction Hamiltonian density (again omitting $\lambda(x)$),
\spliteq{
\mathcal{H}_\mathcal{A}(x) &= -\ii\ \int_{\mathbb{R}^+} \text{d}\omega \sqrt{\tfrac{\omega}{8\pi}} \left((C-\ii S) a_\omega e^{-\ii\omega x} -(C+\ii S)  a^\dagger_\omega e^{\ii\omega x}\right)\\
&= \frac{C}{2}\underbrace{(-\ii)\ \int_{\mathbb{R}^+} \text{d}\omega \sqrt{\tfrac{\omega}{4\pi}} \left(a_\omega e^{-\ii\omega x} -  a^\dagger_\omega e^{\ii\omega x}\right)}_{\Pi_+(x)} - \frac{S}{2}\underbrace{ \int_{\mathbb{R}^+} \text{d}\omega \sqrt{\tfrac{\omega}{4\pi}} \left(a_\omega e^{-\ii\omega x} +  a^\dagger_\omega e^{\ii\omega x}\right)}_{\neq \Phi_+(x)}.
}
The term that goes with $S$ is not an observable that we know of; has a power of $\omega$ more than $\Phi_+$ but not the $\pm\ii$ factors of $\Pi_+$. We can decompose the creation/annihilation operators into field amplitude and momentum operators: 

\spliteq{
a_\omega &= \frac{1}{\sqrt{4\pi}}\int_{\mathbb{R}}\! \text{d}x\, e^{\ii \omega x} \left[ \sqrt{\omega} \Phi_+(x) +\tfrac{\ii}{\sqrt{\omega}}\Pi_+(x)\right],\\ a^\dagger _\omega &= \frac{1}{\sqrt{4\pi}}\int_{\mathbb{R}}\! \text{d}x\, e^{-\ii \omega x} \left[ \sqrt{\omega} \Phi_+(x) -\tfrac{\ii}{\sqrt{\omega}}\Pi_+(x)\right],
}
and plug this back into our expression for our interaction Hamiltonian density 
\spliteq{
\mathcal{H}_\mathcal{A}(x) &= \frac{C}{2}\Pi_+(x) - \frac{S}{2} \int_{\mathbb{R}^+} \text{d}\omega \frac{1}{4\pi} \left(\int_{\mathbb{R}}\! \text{d}y\, \left[ \omega \Phi_+(y) +\ii\Pi_+(y)\right] e^{\ii\omega (x-y)} +  \text{h.c.}\right)\\
&= \frac{C}{2}\Pi_+(x) - \frac{S}{2} \int_{\mathbb{R}}\! \text{d}y\,\frac{1}{4\pi} \left(\left[ \left(\int_{\mathbb{R}^+} \text{d}\omega\,\omega e^{\ii\omega (x-y)}  \right)\Phi_+(y) +\ii \left(\int_{\mathbb{R}^+} \text{d}\omega\,e^{\ii\omega (x-y)}  \right)\Pi_+(y)\right]  +  \text{h.c.}\right)\\
&= \frac{C}{2}\Pi_+(x) - \frac{S}{2} \int_{\mathbb{R}}\! \text{d}y\,\frac{1}{4\pi} \left(\left[ \text{Re} \left(\int_{\mathbb{R}^+} \text{d}\omega\,\omega e^{\ii\omega (x-y)}  \right)\Phi_+(y) -\text{Im} \left(\int_{\mathbb{R}^+} \text{d}\omega\,e^{\ii\omega (x-y)}  \right)\Pi_+(y)\right]  \right)\\
&= \frac{C}{2}\Pi_+(x) + \frac{S}{8\pi} \int_{\mathbb{R}}\! \text{d}y \left( \frac{\Phi_+(y)}{(x-y)^2} +\frac{\Pi_+(y)}{(x-y)} \right).
}
We clearly see that the $S$ term is non-local. Essentially, this is because the $S$ term was ``a factor of $\omega$ off''; note the following integrals
\spliteq{
\int_{\mathbb{R}^+} \text{d}\omega\,e^{\ii\omega (x-y)}   &= \pi \delta(x-y) + \frac{\ii}{(x-y)}\\
\int_{\mathbb{R}^+} \text{d}\omega\,\omega\, e^{\ii\omega (x-y)}   &= - \frac{1}{(x-y)^2} - \ii \pi \delta'(x-y)
}
which is obviously non-local.  

\section{Scaling Inequalities}\label{ApD}
Here we prove the inequality \eqref{eq:3_ineq}. In order to recover locality, we want the teleported teleportation term (second term in \eqref{eq:delta_E_spatial}) to far outweigh the non-locality in $E_A(x)$ (second term \eqref{eq:E_dens_A}) integrated with respect to B's coupling to the field $\mu(x)$, which represents to leading order in distance the energy that B could extract from the non-local part of A's energy injection. To extract a positive amount of work through B's interaction, we need the teleportation term to outweigh the positive energy injected by B from switching costs (we want the second term to outweigh the first term in \eqref{eq:delta_E_spatial}). To both recover locality and have positive work extracted from the field by B, we need both of the following to hold
\spliteq{
&\mu_0^2\! \!\int_{\mathbb{R}}\!\! \text{d}x (F_\textsc{b}(x))^2\ll \text{Im}(\Gamma)\lambda_0\mu_0 \int_{\mathbb{R}^2}\!\!\!\!\text{d}x\, \text{d}y\frac{F_\textsc{a}(x)F_\textsc{b}(y)}{(y-x+T)^2}.\\
&\mu_0\lambda^2_0\int_{\mathbb{R}}\text{d}x\, F_\textsc{b}(x)\left({\int_{\mathbb{R}}\text{d}y\, \frac{F_\textsc{a}(y) }{(y-x+T)^2}}\right)^2 \ll
\text{Im}(\Gamma)\lambda_0\mu_0 \int_{\mathbb{R}^2}\!\!\!\!\text{d}x\, \text{d}y\frac{F_\textsc{a}(x)F_\textsc{b}(y)}{(y-x+T)^2}
}
We can assume for now that $d$ is sufficiently large so that $\text{Im}(\Gamma)\approx1$. Consider $L$ being the distance between support of A and B's coupling functions at time $T$, i.e. the distance between $\text{supp}(F_\textsc{a})$ translated by $T$ to the left and $\text{supp}(F_\textsc{b})$. Then, using $L^p(\mathbb{R})$ norm notation, $\lVert f\rVert_p \equiv (\int_\mathbb{R} |f(x)|^p \text{d}x)^{1/p}$, the above inequalities can be rewritten as 
\spliteq{
&\mu_0^2\lVert F_\textsc{b}\rVert_2^2\ll \lambda_0\mu_0 \frac{\lVert F_\textsc{a}\rVert_1\lVert F_\textsc{b}\rVert_1}{L^2}\\
&\mu_0\lambda^2_0 \frac{\lVert F_\textsc{b}\rVert_1\lVert F_\textsc{a}\rVert_1^2 }{L^4} \ll\lambda_0\mu_0 \frac{\lVert F_\textsc{a}\rVert_1\lVert F_\textsc{b}\rVert_1}{L^2}.
}
Now we can use the fact that we defined $F_\textsc{a}$ and $F_\textsc{b}$ to have unit  $L^1(\mathbb{R})$ norm to write,
\spliteq{
&\mu_0^2\lVert F_\textsc{b}\rVert_2^2\ll \lambda_0\mu_0 \frac{1}{L^2}\\
&\mu_0\lambda^2_0 \frac{1}{L^4}\ll
\lambda_0\mu_0 \frac{1}{L^2}.
}
Cancelling common factors, we can boil these down to
\spliteq{
\mu_0\lVert F_\textsc{b}\rVert_2^2&\ll\frac{ \lambda_0 }{L^2},\\
\frac{\lambda_0 }{L^2}&\ll1,}
which can be combined into
\spliteq{
\mu_0\lVert F_\textsc{b}\rVert_2^2\ll
\frac{\lambda_0 }{L^2}&\ll1.}
Now, by definition, we have $\lVert F_\textsc{b}\rVert_1 =1$, where $F_\textsc{b}$ is of compact support. Suppose we widen the support of this function with respect to a parameter $\sigma>1$ all the while keeping the $L^1(\mathbb{R})$ norm fixed, i.e.
\begin{equation}
F_\textsc{b}(x) \mapsto {F}_B^\sigma(x)\equiv \frac{1}{\sigma}F_\textsc{b}\left(\tfrac{x}{\sigma}\right).
\end{equation}
We can check that this still has a unit 1-norm:
\begin{equation}
\lVert F^\sigma_B\rVert_1 = \int_\mathbb{R} \left|\frac{1}{\sigma}F_\textsc{b}\left(\tfrac{x}{\sigma}\right)\right|\text{d}x = \int_\mathbb{R} \left|F_\textsc{b}\left(\tfrac{x}{\sigma}\right)\right|\frac{\text{d}x}{\sigma}  = \int_\mathbb{R} F_\textsc{b}(u)\text{d}u = \lVert F_\textsc{b}\rVert_1 =1,
\end{equation}
where in the above we performed a integration variable substitution $u\equiv x/\sigma$. We can now check how  $\lVert \hat{F}_B\rVert_2^2$ scales: 
\spliteq{
\lVert F_\textsc{b}^\sigma\rVert_2^2 = \int_\mathbb{R} \left(\frac{1}{\sigma}F_\textsc{b}\left(\tfrac{x}{\sigma}\right)\right)^2\text{d}x = \frac{1}{\sigma}\int_\mathbb{R} F_\textsc{b}^2\left(\tfrac{x}{\sigma}\right)\frac{\text{d}x}{\sigma} = \frac{1}{\sigma}\int_\mathbb{R} F_\textsc{b}^2\left(u\right)\text{d}u = \frac{1}{\sigma}\lVert F_\textsc{b}\rVert_2^2.
}
We clearly see that $\lVert F_\textsc{b}\rVert_2^2$ scales as the inverse of the support stretching parameter $\sigma$. Thus, our double inequality becomes

\spliteq{
\frac{\mu_0}{\sigma}\ll
\frac{\lambda_0 }{L^2}&\ll1.} 
\end{widetext}

\bibliographystyle{apsrev4-1}
\bibliography{Gui,Edu}
\end{document}